# Sampling-Based Attack for Centrality Disruption in Complex Networks.

Fariba Afrin Irany
*University of North Texas*
Texas,USA
faribaafrinirany@my.unt.edu

Soumya Sarakar
*Microsoft India*
West Bengal,India
portkey1996@gmail.com

Animesh Mukherjee
*IIT Kharagpur*
West Bengal,India
animeshm@cse.iitkgp.ernet.in

Sanjukta Bhowmick
*University of North Texas*
Texas,USA
sanjukta.bhowmick@unt.edu

*Abstract*—Many mobile networks are represented as graphs to obtain insight to their connectivity and transmission properties. Among these properties *centrality resilience*, that is, how well centralities, such as closeness and betweennesss, are maintained under attacks is a critical factor for proper functioning of a network. In this paper, *we study the centrality resilience of complex networks by developing attack models to disrupt the rank of the top path-based centrality vertices.*

To develop our attack models, we extend the concept of rich clubs of influential vertices to the more general framework of *scattered rich clubs*. We define scattered rich clubs as dense subgraphs of high centrality vertices that are spread (scattered) across the network. Finding scattered rich clubs, although of polynomial time complexity, is extremely expensive computationally. We use snowball sampling to identify these important substructures as well as to identify which edges to target in our proposed attack models.

Our results over a set of real world networks demonstrate that our proposed algorithm is effective in finding the single or scattered rich clubs efficiently and in successfully disrupting the centrality rankings of the network.

*To summarize*, we propose sampling-based attack models for testing the resilience of networks with respect to centrality rankings. As part of this process, we introduce scattered rich clubs, a generalized form of the rich club model, efficient algorithms to detect them, and demonstrate their relation to network resilience.

## I. INTRODUCTION

In recent years, many mobile networks have been represented as graphs in order to study their connectivity, speed of transmission of information, and resilience under attack [1], [2], [3]. To date, most research has focused on the robustness of networks (or graphs), that is how attacks can disconnect the network [4], [5], [6]. In this paper, we develop attack models to study network resilience with respect to path-based centralities, specifically betweenness and closeness centralities. We term this type of resilience as *centrality resilience*, as opposed to the connectivity resilience of the earlier studies.

**Motivation**: Centrality resilience is more difficult to detect than connectivity resilience. When one part of a network cannot communicate with the rest of the system, it is easy to infer that the cause is due to disconnectivity. Attack on centrality, however, may not disconnect the network, but result in longer distances and more time to transmission when traversing the network. That the increased length of the distances is due to the change in the ranking of the high centrality vertices may not be immediately apparent until the centralities of the system are re-computed.

Therefore, centrality resilience is a very powerful tool for insidiously disrupting the functioning of a mobile network, without a drastic change to its structure. Such techniques can be applied for attacks that are typically malicious (stealth attacks in cybersecurity, where the location of attack cannot be immediately known) as well as benign (reducing the load on bottlenecks, under limited resources without completely disconnecting the network).

Robustness of mobile networks have been studied in context of how centrality can affect the connectivity, stability and robustness [2]. The basic idea is to concentrate attack on nodes with high centrality. However, identifying high centrality is expensive and knowledge of the entire network may not be available to the attacker.

Our work presents a more principled model of attacking nodes present in clusters containing high centrality nodes, and not limited to only a few high centrality nodes. Specifically we develop attack models for altering the top ranked high centrality vertices. More formally, the main problem we tackle in the paper is as follows.

**Objective**: *To develop algorithms for deleting edges (attack models) such that the difference in ranking between the top k ranked high centrality vertices in the original and the changed network is greater than a specified threshold.*

**Key steps and contributions**: Our steps to achieve this objective, and key contributions for each step are as follows.

**Step 1. Identify the structural properties that affect centrality resilience** (Section III). We identify substructures that affect centrality resilience based on the observation of [7] that path-based centralities form dense clusters or "rich clubs" in certain networks, which manifest in the inner cores of a network. *A rich club*, as defined in [7], is an assortative subgraph, where all the vertices have high value of a vertex-based property $p$. In this case, the vertex property is betweeneess or closeness centrality. An immediate application of this observation, which was not discussed in the paper, is that *breaking these rich clubs will affect the ranking of the high centrality vertices.*

The rich club model proposed in [7] was restricted to a

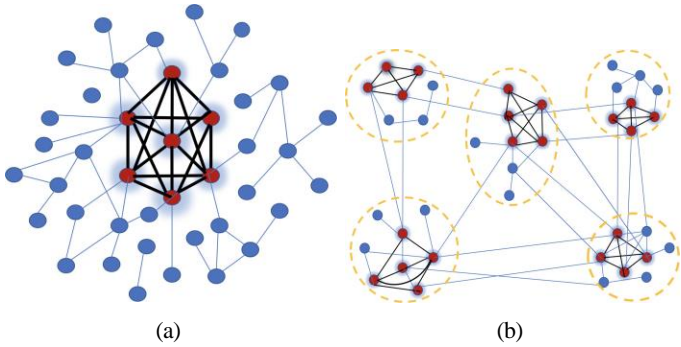

Fig. 1: Two different classes of network. The rich clubs are formed by red vertices. (a) Network with single rich club, dense subgraph is at the core (b) Network with rich clubs scattered across the network.

*single* rich club, formed at the innermost cores of a network. This feature occurs only in specific types of networks. In order to understand centrality resilience of *all types* of networks, we extend the concept of a single rich club to that of *scattered rich clubs*. Scattered rich clubs are clusters that contain high centrality vertices that may be spread across the network. Figure 1 compares the structure of networks with a single (Figure 1 a) and multiple scattered (Figure 1 b) rich club(s).

We use the term "high centrality vertices" to refer to the top-$k$ ranked vertices with high centrality (here betweeneess or closeness). In our experiments we have set $k$ to 20. This means that the union of the vertices in the scattered rich clubs will encompass the top-$k$ ranked high centrality vertices. However, note that, to form dense clusters, some vertices ranked lower than $k$ have to be included as well.

**Key contributions.** We empirically demonstrate that *rich clubs of high centrality vertices can be spread across multiple clusters* of a network. Thus instead of a binary property, the presence of rich clubs is a continuous parameter. We, therefore, develop a metric to measure *the degree of scatteredness for networks*, based on the distribution of high centrality vertices across the clusters.

**Step 2. Develop algorithms to extract the important substructures efficiently** (Section IV). Our next step is to develop algorithms to identify the scattered rich clubs efficiently. To compute the degree of scatteredness, we need to find the high betweenness and closeness centrality nodes as seeds and construct dense clusters around them. However, this approach is very computationally intensive for large networks. In fact, while it is easy to recognize the scattered rich clubs once they are given, it takes several iterative steps to find the appropriate sets of nodes that form the clusters.

We observe that the region surrounding high centrality vertices has expander graph like properties. Specifically, due to their high centrality, the vertices can reach out to many neighbors in a few hops. We use snowball sampling [8] to identify these regions containing high centrality vertices.

**Key contributions.** We develop *an efficient algorithm to find scattered rich clubs using snowball sampling*. We demonstrate that our sampling method can find most of the high centrality nodes and with much lower complexity than the naive method of finding high centrality vertices and then forming clusters.

**Step 3: Leverage the structure of scattered rich clubs to create attack models** (Section V). Our final step is to develop attack models that take into account the structural properties of the rich clubs. Our attacks are based on strategically removing edges from the networks. Attacks based on removing the vertices are equivalent to removing multiple edges. We therefore posit that edge removal is a more insidious and fine grained operation where the attack is spread strategically across the networks, rather than being concentrated at certain vertex points.

In real life an attacker may not have knowledge of the whole network. Therefore, a realistic attack model should be based on sampling. Leveraging our work in Step 2, we use snowball sampling to find the single/scattered rich clubs, and then select edges to delete from these rich clubs.

**Key contributions.** We develop *sample-based attack strategies for disrupting the rank of high centrality vertices*, based on single and scattered rich clubs. We compute the effectiveness of our work by how much the ranking of the vertices have been perturbed, as per the Jaccard index. Our results show that our attack models are indeed effective for networks with both single and scattered rich clubs.

## II. DATASETS AND DEFINITIONS

We provide the definitions of the terms used in this paper and describe the networks that we used for our experiments.

### A. Definitions of properties

We provide the definitions of few key concepts used in this paper, as follows.

*Definition 1:* **Closeness centrality** of a vertex $v$ is the average of the shortest distance between that vertex and all other vertices in the network. It is calculated as $CC(v) = \sum_{s \neq v \in V} \frac{1}{dist(v,s)}$, where $dist(v, s)$ is the length of the shortest path between $v$ and $s$.

*Definition 2:* **Betweenness centrality** of a vertex $v$ is the ratio of the number of shortest paths between a vertex pair that passes through $v$ and all the shortest paths possible between that pair. It is given by $BC(v) = \sum_{s \neq v \neq t \in V} \frac{\sigma_{st}(v)}{\sigma_{st}}$, where $\sigma_{st}$ is the total number of shortest paths between $s$ and $t$, and $\sigma_{st}(v)$ is the total number of shortest paths between $s$ and $t$ that pass through $v$.

*Definition 3:* **$k$-core**: For a graph $G(V, E)$, where $V, E$ are the set of vertices and edges respectively, a $k$-core or network degeneracy is a maximal set of nodes denoted by $C_{\delta_{max}}$ such that each node in $C_{\delta_{max}}$ has at least $\delta_{max}$ neighbors.

The core number of a node is the highest value $k$ such that the node is a part of a $k$-core. The $k$-core decomposition is the assignment of core numbers to nodes.

*Definition 4:* **Rich club subgraph**: Given a graph $G(V, E)$, a vertex-based property, $f$, and a threshold value, $p$, a *rich*

*club* is an assortative subgraph, $S(V_S, E_S)$, $V_S \in V$ and $E_S \in E$, where for all $v \in V_S$, $f(v) \geq p$.

All vertices belonging to a rich club will have at least value *p* of the specified property. Typically rich club is defined for degree and for a certain minimum size[1]. If *p* is set to the highest degree, it is easy to see that the innermost core of a network corresponds to a rich club. Sarkar et al. [7] observed that innermost cores can also form rich clubs with respect to path-based centrality metrics. Generally, a sufficiently high value of *p* is selected such that a single rich club is formed.

## B. Description of test networks

We give a brief description of each category of networks used in our test suite here. These networks are collected from website SNAP(Stanford Large Network Dataset) [9] and Network repository [10]. A summary of the properties of the networks is given in Table I.

- **Autonomous system:** (as19981122, as20000102, as-caida20071105, oregon-1, oregon-2). The network is formed of routers comprising the Internet is organized into subgraphs called Autonomous Systems (AS). Each AS exchanges traffic flows with its neighbors (peers).
- **Peer to peer network:** (p2p-Gnutella24). This is a sequence of snapshots of the Gnutella peer-to-peer file sharing network from August 2002. Nodes represent hosts in the Gnutella network topology and edges represent connections between the Gnutella hosts.
- **Citation network:** (cit-HepPh, cit-HepTh, cora). These are citation networks where papers are represented as nodes and citation links are represented as edges.
- **Collaboration network:** (ca-AstroPh, ca-Condmat). These networks are formed by authorship relationship between authors. Nodes are authors, and they are connected if the authors co-authored papers together.
- **Biological network:** (bio-dmela, bio-grid-fission-yeast) These are protein interaction networks where nodes are proteins and chemical reactions represent the edges.
- **Social network:** (email-enron, email-univ, soc-youtube, facebook). These networks have been constructed from the interaction between users. Here users are represented as nodes and interaction in terms of email exchange or social exchange are represented as edges.
- **Infrastructure network:** (inf-power, inf-openflight, inf-euroroad, california). Here inf-power is a network created from the US power grid. inf-openflight contains ties between two non-US-based airports, inf-euroroad contains connection between two roads located mainly in Europe and california contains connections between roads of california.
- **wikipedia network:** (wiki-vote). Wiki-vote is a wikipedia network on who-votes-on-whom.

---

[1] https://en.wikipedia.org/wiki/Rich-club_coefficient

| Network | Nodes | Edges | Avg Clus Co-eff | Max Core |
|---|---|---|---|---|
| as20000102 | 6474 | 12572 | 0.25 | 12 |
| as-caida | 26475 | 53381 | 0.20 | 22 |
| bio-dmela | 7393 | 25569 | 0.01 | 11 |
| inf-power | 4941 | 6594 | 0.08 | 5 |
| p2p-Gnutella24 | 26518 | 65369 | 0.01 | 5 |
| ca-AstroPh | 18772 | 198050 | 0.63 | 56 |
| ca-CondMat | 23133 | 93439 | 0.63 | 25 |
| cit-HepPh | 34546 | 420877 | 0.29 | 30 |
| cit-HepTh | 27770 | 352285 | 0.31 | 37 |
| cora | 23166 | 89157 | 0.26 | 13 |
| email-enron | 35692 | 183831 | 0.49 | 43 |
| email-univ | 1100 | 5500 | 0.22 | 12 |
| bio-grid | 2000 | 25300 | 0.19 | 69 |
| california | 9700 | 16200 | 0.05 | 15 |
| inf-euroraod | 1200 | 1400 | 0.01 | 3 |
| inf-openflight | 2900 | 30500 | 0.40 | 56 |
| oregon-1 | 11500 | 23400 | 0.28 | 18 |
| oregon-2 | 11800 | 32700 | 0.34 | 32 |
| facebook | 3900 | 137600 | 0.26 | 57 |
| wiki-vote | 889 | 2900 | 0.15 | 10 |

TABLE I: Test suite of networks; along with their size, average clustering co-efficient, and largest core number.

## III. SCATTERED RICH CLUBS

We empirically show that high centrality vertices can be distributed across the network cores and propose a new metric, the *degree of scatteredness*, to quantify this distribution.

## A. Distribution of rich clubs of high centrality vertices

It was reported in [7], that in several networks, the high betweeness and closeness centrality vertices were located in the innermost cores. Since the innermost cores form a dense subgraph, therefore this becomes a rich club. However, as seen in Figure 2, in many networks the high centrality vertices can be distributed across cores. This phenomena requires a more general definition, that of *scattered rich clubs*.

*Definition 5:* **Scattered rich club**: Given a graph $G(V, E)$, a vertex-based property $f$, and a threshold value, *p*, *scattered rich clubs* are a *set of disjoint* assortative subgraphs, $\{S_1, S_2, \ldots, S_n\}$, where $S_i(V_i, E_i)$, such that $V_1 \cap V_2 \cap \ldots \cap V_n = \phi$ and $\forall v \in (V_1 \cup V_2 \cup \ldots \cup V_n), f(v) \geq p$.

Similar to rich clubs, the membership to scattered rich clubs is also determined by the value of *p*. However, the implicit expectation of a single subgraph is relaxed. This simple modification, generalizes the definition of rich clubs, because every network will have scattered rich clubs, even if the number of vertices in the rich clubs is one.

Thus there are two extremes; on one end is the single rich club, which is a cluster containing all the high centrality vertices located at the innermost core, and on the other end is a network where the high centrality vertices are not connected and each form a rich club of size one. Note that if the rich club sizes are very small, the cluster will not be large enough to develop attack models that alter the ranking of high centrality vertices without disconnecting the network. Thus, in the practical context, we further relax the constraints of scattered rich clubs to obtain clusters that contain the high centrality vertices and are also "large enough", i.e., greater than a specified threshold.

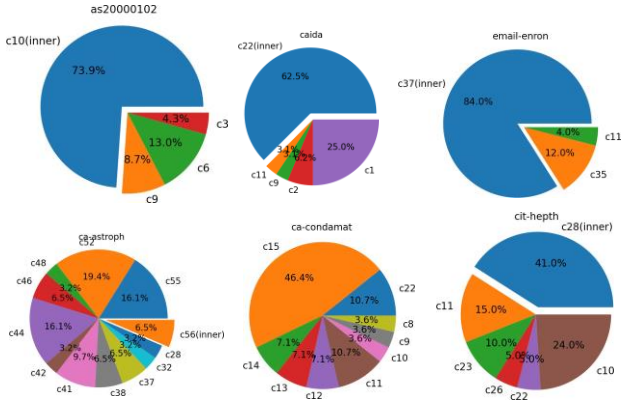

Fig. 2: Pie charts showing the distribution of the top 20 high centrality nodes (union of the set of top 20 high betweenness centrality and top 20 high closeness centrality nodes) across the cores of the networks. The top panel shows examples of networks that tend towards a single rich club. Most high centrality vertices are in the top cores. The bottom panel shows examples of networks that have scattered rich clubs.

*To summarize*, each (scattered) rich club should contain at least one high centrality node, but they can be augmented with a few neighboring nodes to produce a sufficiently dense and non-trivial cluster.

### B. Identifying clusters forming the scattered rich clubs

We explore how to identify these scattered rich clubs in the network. The work of Estrada *et. al.* [11] shows that scale-free real world networks, such as the ones in Table II, fall in two categories. Either it has a dense core with sparsely connected periphery (single rich club) or the network is composed of modular units which are individually densely connected but have sparse inter module connection (scattered rich club).

We identify scattered rich clubs using the following steps (also illustrated in Figure 1;

*First*, we obtain the top k high betweenness ($N_{hbc}$) and closeness ($N_{hcc}$) centrality vertices of the network. Here we set k=20. We take the union of the two sets to create a unified set of high centrality vertices $N_{hc} = N_{hcc} \cup N_{hbc}$.

*Second*, we form a cluster comprising of each node in $N_{hc}$ and its neighbors. The total number of clusters formed will be equal to the number of high centrality nodes ($|N_{hc}|$).

*Third*, we merge overlapping clusters. For each pair of clusters $C_i$, $C_j$, we take the intersection of $C_i$ and $C_j$ to identify the common elements between the two sets. If there are common nodes, then we merge the two clusters $C_i$ and $C_j$, such that $C_i = C_i \cup C_j$. We iterate the merging step until no more clusters can be merged.

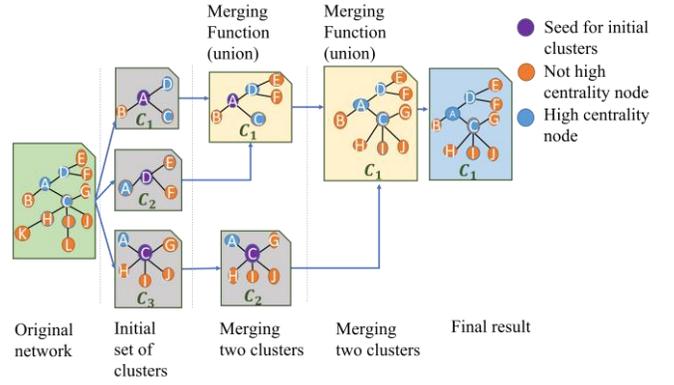

Fig. 3: Step by step illustration of the clustering algorithm

It is easy to see that at the end of these steps, we will have the disjoint clusters containing the high centrality vertices and their neighbors. Note the neighbors of a high centrality node generally also have medium to high centrality. If the network has a single rich club, then there will be one cluster, otherwise there will be multiple clusters.

### C. Quantifying the degree of scatteredness

We observe that for scattered rich clubs, the number of clusters, and the number of high centrality nodes in each cluster vary. We quantify this distribution as follows;

Let the total number of high centrality vertices be $H$ and the total number of clusters be $K$.

Let the total number of high centrality nodes in cluster $C_i$ be $H_i$. We order the clusters such that if $i < j$, $H_i \geq H_j$ ($1 \leq i \leq K$); i.e. in the decreasing number of high centrality vertices in them. We then compute the ratio of the number of high centrality vertices in the cluster to the number of high centrality vertices not yet seen. Formally, if the total number of high centrality vertices is $H$, then the ratio for cluster $C_x$ is $R_i = \frac{H_x}{H - (\sum_{i=1}^{i=x-1} H_i)}$.

The degree of scatteredness is computed as the geometric mean of the ratios;

$$(\Pi_{i=0}^{i=K} R_i)^{1/K}$$

From this formula it can be seen, that a single cluster will give the degree of scatterdness 1, which is the maximum value. When every cluster has just one vertex, the value will be $(1/K!)^{1/K}$, which will tend to zero as $K$ becomes large. Table III gives the degree of scatteredness for each network. As can be seen, the more the high centrality nodes scatter into clusters, the value of scatteredness becomes lower.

## IV. IDENTIFYING RICH CLUBS USING SNOWBALL SAMPLING

We now present an efficient algorithm to identify the scattered rich clubs in a given network. While the steps presented in Section III can locate these rich clubs, in practice, the process of computing the high centrality vertices for large networks can be very computationally intensive. Moreover,

| Dataset | Number of Clusters | Distribution of High Centrality Nodes | Degree of Scatteredness |
|---|---|---|---|
| as20000102 | 1 | 24 | 1 |
| caida | 1 | 40 | 1 |
| email-enron | 1 | 25 | 1 |
| inf-openflight | 1 | 27 | 1 |
| oregon-1 | 1 | 23 | 1 |
| oregon-2 | 1 | 23 | 1 |
| wiki-vote | 1 | 31 | 1 |
| email-univ | 1 | 26 | 1 |
| bio-dmela | 2 | 24,1 | .98 |
| ca-astroph | 2 | 29,2 | .97 |
| cit-hepth | 2 | 29,11 | .85 |
| california | 3 | 31,5,1 | .88 |
| inf-power | 3 | 20,6,3, | .77 |
| facebook | 4 | 37, 1(3) | .63 |
| bio-grid-fission-yeast | 4 | 30, 1(3) | .62 |
| inf-euroroad | 6 | 24,4(2), 2(2) | .55 |
| cora | 6 | 20,4, 1(4) | .49 |
| ca-condamat | 4 | 23,2, 1(2) | .47 |
| cit-hepph | 12 | 18,7,5,2, 1(8) | .31 |
| p2p-gnutella24 | 26 | 3, 2(8), 1(17) | .11 |

TABLE II: The degree of scatterredness of the networks. The distribution of high centrality nodes ($N_{hc}$) per cluster is listed. For multiple clusters with same number of high centrality vertices, the multiplicity is given in parenthesis, i.e. 1(3), means there are three clusters with 1 high centrality vertex. Blue: networks with single rich club. Yellow: Networks with scattered rich clubs.

in real world applications, the entire network may not be available for analysis.

To address these challenges, we propose identifying the rich clubs by sampling the network. To select the sampling algorithm, we observe that the rich clubs exhibit characters similar to *expander graphs*.

Expander graphs are graphs that are sparse, but the vertices are highly connected [12]. Given a graph $G(V, E)$, let $S \subset V$, and $N(S)$ be the set of neighbors of $S$ that are in $V - S$. The *maximum expansion factor* for a set of size $k$, is given by $X(S) = argmax_{S:|S|=k} \frac{|N(S)|}{|S|}$. A graph is a $(k, \alpha)$ expander if $|N(S)| \geq \alpha |S|$ for each $S \subset V$, and $|S| \leq k$ [8].

*Snowball sampling* was presented in [8], [13], where the authors conjectured that samples with higher expansion factors are more likely to be representative of the community structure of the network. A brief description of snowball sampling algorithm is as follows; *(i)* Initialize the set of sample vertices, $S$, to a null set. *(ii)* Start with a seed vertex, $v$, that is added to the $S$, *(iii)* Based on the neighbors of $S$, $N(S)$, select a new vertex $v$, such that $|N(v) - (N(S) \cup S)|$ is maximized. That is, select a new vertex such that the number of new neighbors added to $N(S)$ is maximized. *(iv)* Add $v$ to $S$. *(v)* Continue the expansion, until $|S|$ is larger than a given threshold, $k$.

We posit that the rich clubs are good expanders, since the high centrality vertices embedded in them can, in a few hops, reach a wide set of vertices. Based on this hypothesis, we aim to use snowball sampling to find the high centrality vertices.

While our algorithm follows the basic steps of snowball sampling, some of the unique features are as follows;

TABLE III: Prediction of high centrality nodes by snowball sampling. HCN is the number of high centrality nodes in the actual network. Cluster is the number of clusters found using sampling. The actual number is given in the parenthesis. The last two columns give the precision and recall values of the predicted high centrality nodes. Networks with recall values less than .70 are marked with a star. Blue: networks with single rich club. Yellow: Networks with scattered rich clubs.

| Dataset | HCN | Seed Node | Cluster | Precision | Recall |
|---|---|---|---|---|---|
| **as2000102** | 24 | Random | 1 | .6 | 1 |
| | | HD+HCC | 1 | .6 | 1 |
| **caida** | 40 | Random | 1 | .87 | .87 |
| | | HD+HCC | 1 | .8 | .8 |
| **email-enron** | 25 | Random | 1 | .62 | 1 |
| | | HD+HCC | 1 | .62 | 1 |
| **inf-openflight** | 27 | Random | 1 | .62 | .92 |
| | | HD+HCC | 1 | .62 | .92 |
| **oregon-1** | 23 | Random | 1 | .57 | 1 |
| | | HD+HCC | 1 | .57 | 1 |
| **oregon-2** | 23 | Random | 1 | .57 | 1 |
| | | HD+HCC | 1 | .57 | 1 |
| **wiki-vote** | 31 | Random | 1 | .77 | 1 |
| | | HD+HCC | 1 | .77 | 1 |
| **email-univ** | 26 | Random | 1 | .55 | .85 |
| | | HD+HCC | 1 | .55 | .85 |
| **bio-dmela** | 24 | Random | 1(2) | .55 | .88 |
| | | HD+HCC | 1(2) | .62 | 1 |
| **ca-astroph** | 31 | Random | 1(2) | .77 | 1 |
| | | HD+HCC | 1(2) | .77 | 1 |
| **cit-hepth *** | 40 | Random | 1(2) | .62 | .62 |
| | | HD+HCC | 1(2) | .6 | .6 |
| **california** | 33 | Random | 3(3) | .65 | .79 |
| | | HD+HCC | 3(3) | .7 | .85 |
| **inf-power*** | 29 | Random | 1(3) | .08 | .10 |
| | | HD+HCC | 1(3) | .15 | .20 |
| **facebook** | 32 | Random | 1(4) | .6 | .75 |
| | | HD+HCC | 1(4) | .57 | .72 |
| **bio-grid-fission-yeast** | 33 | Random | 3(4) | .65 | .78 |
| | | HD+HCC | 3(4) | .7 | .84 |
| **inf-euroroad*** | 33 | Random | 2(6) | .05 | .06 |
| | | HD+HCC | 2(6) | .05 | .06 |
| **cora** | 28 | Random | 6(6) | .55 | .79 |
| | | HD+HCC | 6(6) | .52 | .75 |
| **ca-condamat** | 28 | Random | 3 4) | .7 | 1 |
| | | HD+HCC | 3(4) | .7 | 1 |
| **cit-hepph** | 40 | Random | 12(12) | 1 | 1 |
| | | HD+HCC | 12(12) | 1 | 1 |
| **p2p-gnutella24** | 36 | Random | 24(26) | .8 | .88 |
| | | HD+HCC | 23(26) | .75 | .83 |

1) We experimented with two methods of selecting the seed nodes. A random node, as is the default, a node that has both high degree and high clustering co-efficient.
2) We set the threshold, $k$, as 10% of the nodes in the network.
3) Each run of snowball sampling yields only one expander graph. Since our goal is to find all the scattered rich clubs, we run the snowball sampling multiple times.
4) After each run of the sampling we obtain a sampled subgraph. We analyze the core periphery structure of this subgraph, and designate the nodes in the innermost and second innermost core as high centrality nodes.
5) We continue obtaining new snowball samples, until the set of high centrality nodes do not change, or a maximum

threshold of runs (here set to 40) have been executed.

At the end of this process, the sampled subgraphs are the *predicted rich clubs* and the nodes in their innermost and second innermost cores are the *predicted high centrality vertices.*

The *complexity* of this algorithm is dominated by the size of the subgraph $S$. During the snowball sampling, each edge in the subgraph is accessed once, as new nodes are sampled. During computing the core periphery, again each edge in the subgraph is accessed once, because core periphery computation is linear time. Thus the total complexity per sampled graph is bounded by the number of edges in $S$, say $E(S)$. The total complexity will be $O(T * E(S))$, where $T$ is the maximum number of iterations. In contrast, the complexity for both high betweenness and high closeness centrality vertices is $O(V(V+E))$–an order of magnitude higher.

### A. Results and discussion of the sampling algorithm

Table III shows the effectiveness of the snowball sampling in finding high centrality vertices. We experimented with two methods of selecting seed nodes, (i) using a random seed and (ii) using a seed that has high degree and high clustering coefficient. We extended each subgraph until 10% of the nodes in the network were covered. We create the ground truth, by taking the union of the set of top 20 high betweenness centrality vertices and the top 20 high closeness centrality vertices. Because some nodes can be common among the two sets, the cardinality of the union can be less than 40.

We compute the core-periphery structure of each sampled subgraph, and assign the nodes in the innermost core and second innermost core in a set of probable high centrality ones. After the requisite number of samples have been observed, we take the top 40 high degree nodes from the set of probable high centrality vertices to form the predicted set of high centrality nodes. Since we do not know the size of the ground truth or the number of actual clusters of high centrality nodes in the complete network, in our sampling we always select a set of 40 high centrality nodes. We compute the precision and recall of these predicted nodes as compared to the ground truth.

As seen in Table III, the precision values of our predicted set is generally low. This is because the total predicted set size, 40, can be higher than the ground truth. However, the recall, whether all the nodes in the ground truth were obtained is 80% or higher for all the networks with a single rich club (rows colored blue). The recall is also high, more that .70, for most of the networks with scattered rich clubs. We also note that the sampling algorithm in general finds fewer clusters than the actual ones given in Table II. The recall will still remain high, so long as the main cluster, containing most of the high centrality nodes is part of the sampled subgraph.

The networks with the lowest recall are inf-power and inf-euroroad. This is because these networks are nearly planar, and therefore do not have any tight clusters. As a result, snowball sampling, which is designed to find dense regions in the network, cannot correctly identify the region with high centrality nodes. The network cit-hepth also has low recall.

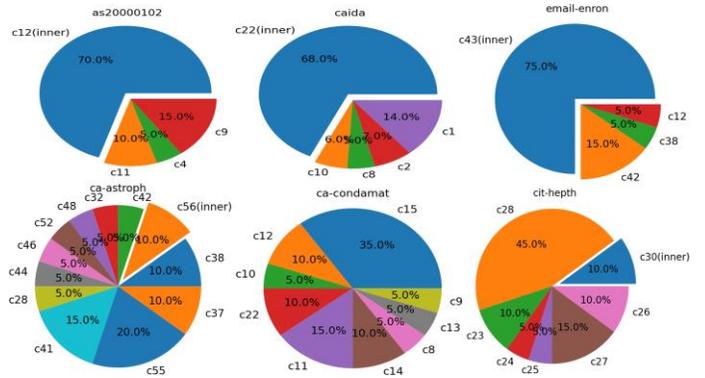

Fig. 4: Distribution of high centrality nodes distribution across different cores for subgraph obtained using 10 percent node sampling by snowball algorithm, with nodes with (High degree and high clustering coefficient node used as initial sample).Top panel are the networks with a single rich club. Bottom are the networks with scattered rich clubs.

This is because the sampling algorithm only found one cluster. The second cluster that had about one fourth of the high centrality nodes was not found.

### V. ATTACK MODELS FOR DISRUPTING CENTRALITY

We present our attack model for selecting edges to remove from the network, such that the ranking of the high centrality nodes is disrupted. As seen in Section IV, snowball based sampling is effective in finding the high centrality vertices in many of the networks with single or scattered rich clubs. As further seen in Figure 4, even with one instance of sampling, the distribution of high centrality vertices in the core is roughly equivalent to their distribution in the original network. Based on these observations, we develop an attack model based on a similar snowball sampling.

**Sampling based attack.** The main objective of the attack model is to identify the edges to be removed from the network. To do so, we first run one instance of subgraph sampling, taking a vertex with high degree and high clustering coefficient as seed, such that $\approx 10\%$ of the nodes in the network are in the sample. We select the the edges to be removed from this sampled network using the following criteria. In the *first criterion*, we select an edge if at least one of its end points has a high core number in the sampled graph. This means that the connection of one possible high centrality vertex is being removed. In the *second criterion*, we select an edge if both its endpoints have a high core number in the sampled graph. We deem a node to have high core number if it is in the inner or second innermost core of the sample graph.

We test the centrality resilience by removing 2%, 4%, 6% and 8% edges. The percentages are set with respect to the total network. As we go to higher percentages, there may not be enough edges to remove that follow the edge selection criteria. For the first criterion, if we reach a limit of edges to choose, we stop the process. For the second criterion, if we reach a limit on the edges, we then select edges using the first criterion.

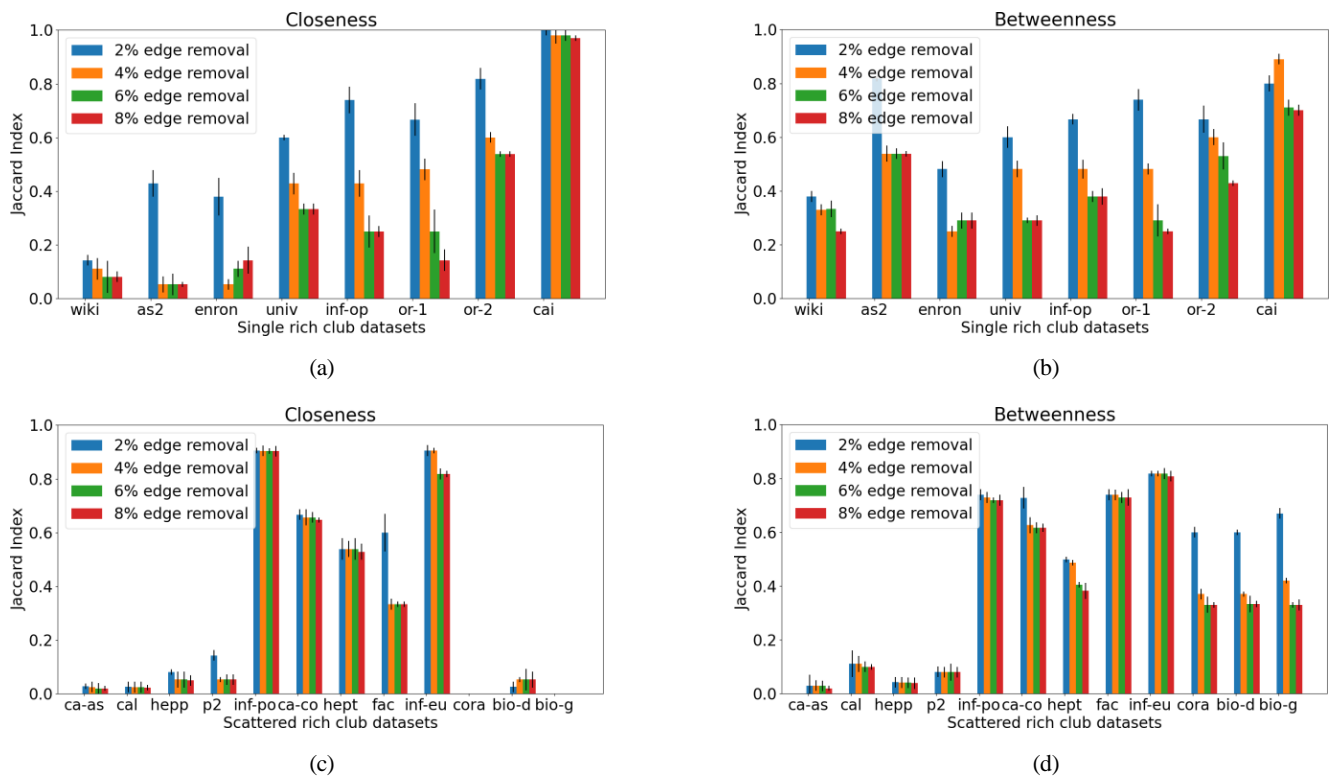

Fig. 5: Results for attack model using the first criterion and high degree and high clustering co-efficient node as the initial seed. (a) single rich club network, closeness centrality (b) single rich club network, betweenness centrality centrality, (c) scattered rich club network, closeness centrality (d) scattered rich club network, betweenness centrality.

After each set of edges is removed we identify the top 20 high betweenness and closeness centrality vertices. We compare these high centrality vertices in the perturbed network with the ones in the original network using the *Jaccard index*. Jaccard index measures the similarity between the two sets $A$ and $B$, as $J(A, B) = \frac{A \cap B}{A \cup B}$ The value can range from 0 to 1. The closer the value is to the more perturbed the network is due to the attack.

## VI. Results of the Attack Models

We show using the Jaccard index how the ranking of the top 20 closeness and betweenness centrality vertices changed under the attack in Figure 7 (the first criteria) and Figure 8 (the second criteria). We experimented with the attack 5 times. The histograms report the average values and the standard deviation of the Jaccard Index.

While Jaccard index decreases in most cases, in both networks we see a larger decrease for the networks with scattered rich clubs as compared to the networks with a single rich club. The single rich club network structure is more difficult to perturb as the high centrality nodes tend to cluster together in the innermost core of the network. On the other hand, it is easier to perturb the scattered rich club network structure, as multiple rich clubs are available, and are smaller in size. We also observe that closeness centrality rankings are more affected, have lower Jaccard index, by the attack as compared to the betweenness centrality.

Networks that has higher recall as per Table III are more affected by the attack model. This is because in these networks, the snowball sampling is more likely to find the correct high centrality nodes. Comparing Figures 7 and 8 we see that the second criterion impacts single rich clubs more than scattered ones. We hypothesise that this is because the stricter criteria makes it more likely that edges within the rich club get selected.

## VII. Related Work

We review some of the pertinent research on attack models and network resilience. **Rich clubs**: Emergence of rich club or *core-tier* was first empirically shown by [14]. Measures to detect rich clubs are based on assortativity [15] and random walks [16]. There are also studies on how rich clubs affect the overall network dynamics such as higher order clustering [17], dyadic effect [18] and cortical transmission [19]. To the best of our knowledge this is the first work which proposes that rich clubs can be scattered across network and empirically demonstrates the relation between network resilience and rich club organization.

**Core-periphery structure**: Coreness information introduced by Seidman et. al. [20] is useful in network mining tasks such as community detection [21], information propagation [22],

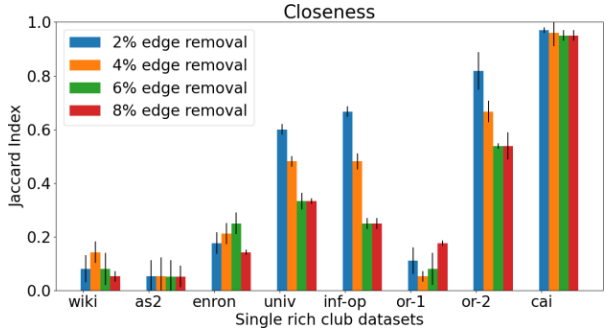
(a)
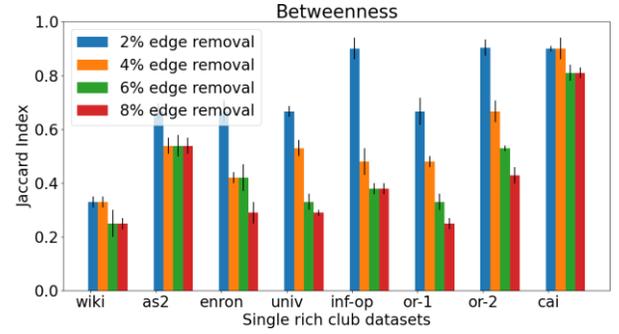
(b)
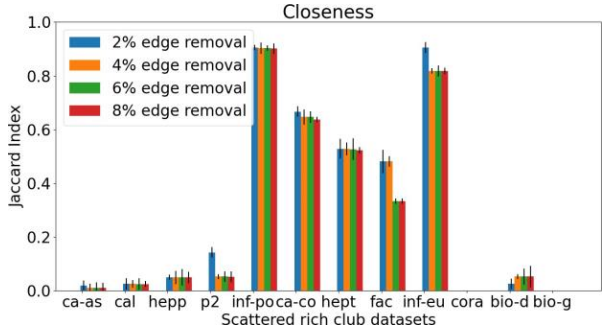
(c)
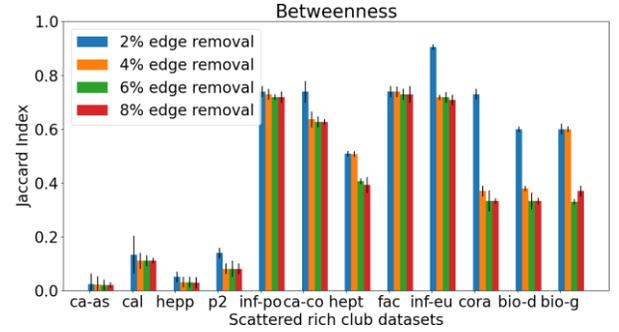
(d)

Fig. 6: Results for attack model using the second criterion and high degree and high clustering co-efficient node as the initial seed. (a) single rich club network, closeness centrality (b) single rich club network, betweenness centrality centrality, (c) scattered rich club network, closeness centrality (d) scattered rich club network, betweenness centrality.

anomaly detection [23]. Several works have focused on efficient algorithms to compute coreness [24], [25], [26].

**Network resilience**: Understanding network structure that governs robustness is an important research area especially after seminal works of Barabási *et. al.* [4]. For recent works on network robustness we refer the reader to [5], [6]. Adiga *et. al.* [27] studied the robustness of the top cores under sampling and in noisy networks and Laishram *et. al.* [28] developed core based approaches to increase network resilience. Here, we focus on resilience with respect to centrality and evaluate this property under random, core based and rich club based attacks.

**Attack Models**: There have been several papers on attack models that perturb the centrality of top k high centrality nodes. The attack models proposed in [29] target nodes with high importance in a large scale free network, and the average time complexity of the three models proposed in the paper is $O(m)$. A heuristic algorithm is presented in [30] to balance the centrality measures in a network by link addition. In [31] authors propose an attack model targeting nodes with high eigenvector centrality. The authors in [32] investigated the effect on network structure when nodes are targeted based on the non-local measure of properties such as degree distributions, clustering coefficients, and assortativity coefficients .

An exploration in the area of the structural centrality-based targeted attack is the main goal in [33], and the paper also provided a comparative analysis of their result with the degree-based targeted attack. To check the robustness of the chinese power grid system, [34] presented different attack strategies and proposed that increasing the load capacity of network edge can increase the robustness of the network. In [35] a disruptive energy measure to target a network was proposed along with a new measure of centrality based on the gravity law, and it was shown that the targeted attack following the proposed measure is more efficient than the other measures for specific values of the parameter. The authors in [36] concluded that centrality(degree and betweenness) based attack strategies become more effective when node centrality is updated after each removal step after computing the percolation thresholds.

Attack models have also been developed for specific applications. Models of attack to economies in international oil trade network(iOTN) were presented in [37] where nodes with the higher importance rankings(high degree centrality, high local clustering coefficient) were chosen for removal. A new ensemble learning-based critical node removal attack (ECNRA) algorithm was proposed in [38] where the authors showed that damage of the robustness using the proposed algorithm is worse compared to the degree-based attack and random node attack. Another attack model based on node removal was presented in [39] where USA flights considering the time were scheduled as a temporal network, and top-ranked airports(high degree centrality, high closeness centrality, high betweenness centrality) are chosen as target attack nodes. An attack model based on high centrality node removal is used to analyze robustness into an inter-cities mobility network in [40] wherewith the abstraction of municipal initiatives is considered as failures of nodes, and the federal actions are thought of as targeted attacks for the network. Several other works [41][42][43][44] present techniques for removing nodes and their associated edges from the network.

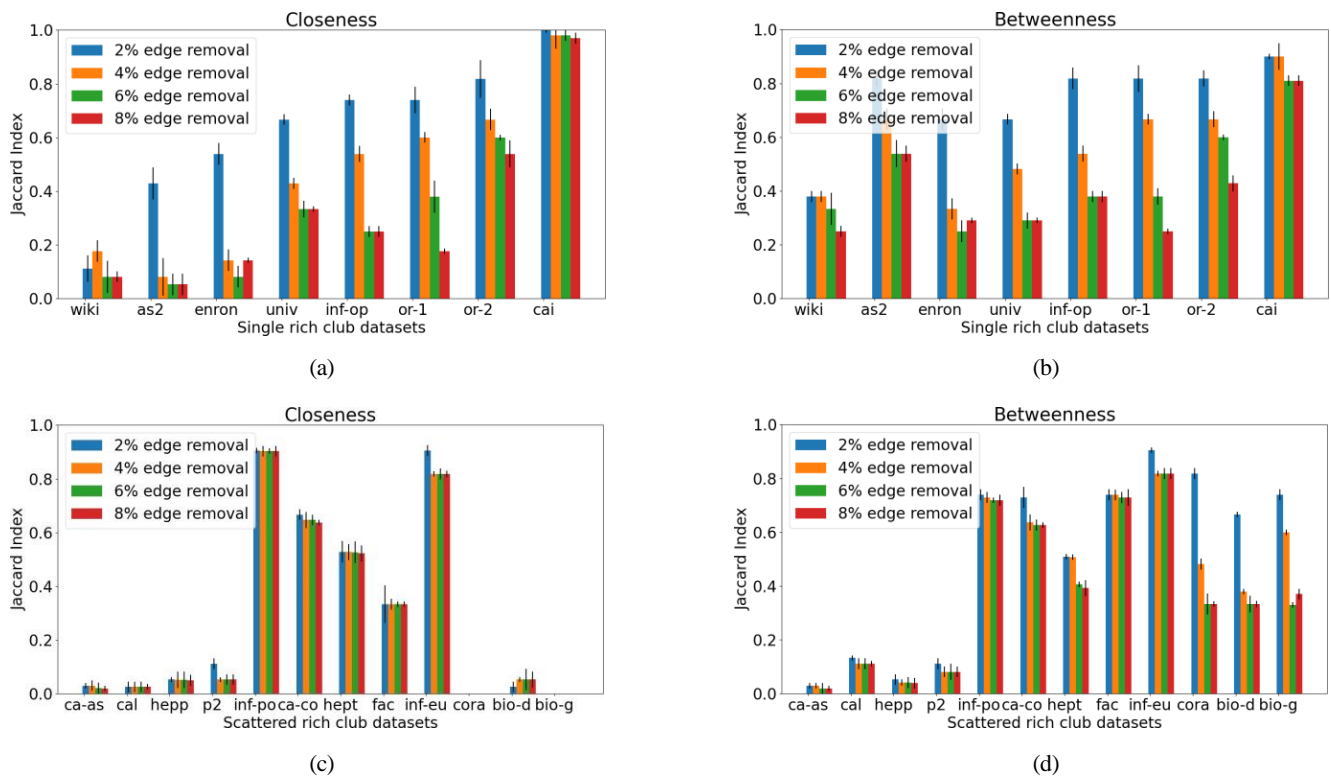

Fig. 7: Results for attack model using the first criterion and random node as the initial seed. (a) single rich club network, closeness centrality (b) single rich club network, betweenness centrality centrality, (c) scattered rich club network, closeness centrality (d) scattered rich club network, betweenness centrality.

## VIII. CONCLUSIONS AND FUTURE WORK

In this work we, for the first time, propose the concept of scattered rich clubs and show that rich club of central nodes can be scattered across the network as opposed to being concentrated at the core. We discuss the implications of this organization in terms of network centrality resilience and develop a predictive approach for discovering the scattered rich clubs using snowball sampling. Our main conclusions are as follows.

We believe that scattered rich club can provide more detailed insights to how change occurs in complex networks. In future, we plan to study how scattered rich clubs affect other network properties such as communities and whether they can be used to predict high centrality nodes in dynamic networks.

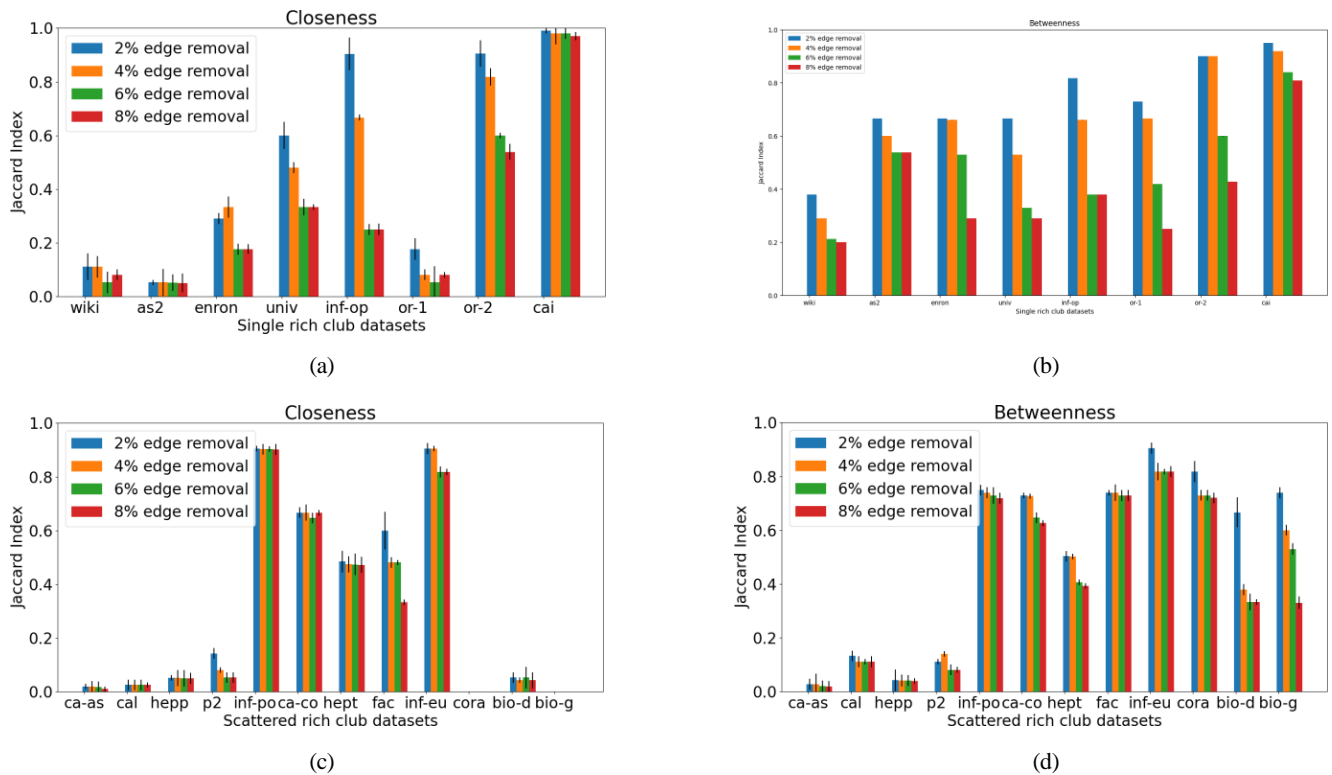

Fig. 8: Results for attack model using the second criterion and high degree and random node as the initial seed. (a) single rich club network, closeness centrality (b) single rich club network, betweenness centrality centrality, (c) scattered rich club network, closeness centrality (d) scattered rich club network, betweenness centrality.